\documentclass[%
 reprint,%
 amssymb, amsmath,%
 aip,cha,%
]{revtex4-1}

\usepackage{epsfig}
\usepackage{amsmath}
\usepackage{color}
\def\notes#1{ } 
\def\be{\begin{equation}}
\def\ee{\end{equation}}
\def\bea{\begin{eqnarray}}
\def\eea{\end{eqnarray}}

\def\bra #1 {\langle {#1} \vert}
\def\ket #1 {\vert {#1} \rangle}

\begin{document}
\renewcommand{\thefootnote}{\fnsymbol{footnote}}

\title{Multi-state effective Hamiltonian and size-consistency corrections in stochastic configuration interactions}

\author{Seiichiro L. Ten-no}
\email[]{E-mail: tenno@garnet.kobe-u.ac.jp}

\affiliation{Graduate School of Science, Technology, and Innovation, Kobe University, Nada-ku, Kobe 657-8501, Japan}

\date{\today}

\begin{abstract}
Model space quantum Monte Carlo (MSQMC) is an extension of full configuration interaction QMC (FCIQMC) that allows us to calculate quasi-degenerate and excited electronic states by sampling the effective Hamiltonian in the model space.
We introduce a novel algorithm based on the state-selective partitioning for the effective Hamiltonian using left eigenvectors to calculate several electronic states simultaneously at much less computational cost than the original MSQMC with the energy dependent partitioning.
The sampling of walkers in MSQMC is analyzed in the single reference limit using a stochastic algorithm for higher-order perturbation energies by the analogy of the deterministic case utilizing a full configuration interaction program.
We further develop size-consistency corrections of the initiator adaptation (i-MSQMC) in three different ways, {\it i.e.} the coupled electron pair approximation, a posteriori, and second-order pertrubative corrections. It is clearly demonstrated that most of the initiator error is originating from the deficiency of proper scaling of correlation energy due to its truncated CI nature of the initiator approximation, and that the greater part of the error can be recovered by the size-consistency corrections developed in this work.
\end{abstract}

\maketitle
\section{INTRODUCTION}
Quantum Monte Carlo (QMC) in configuration space has become an important choice of quantum chemical tools for accurate electronic structures in recent years.\cite{afqmc03,afqmc07,fciqmc09,i-fciqmc,sqmc}
The vast majority of such QMC approaches are based on the projection operator ansatz,
\be
\psi(\tau)=e^{-\tau(\hat H-E)}\psi(0), \label{eq:proj}
\ee
that converges to the ground state by projecting out the excited state components repeatedly from $\psi(0)$ having nonzero overlap to the ground state wavefunction.
Especially, the full configuration quantum Monte Carlo (FCIQMC) advocated by Alavi, Booth and coworkers is one of the important means that can handle very large Hilbert space problems in near FCI accuracy.\cite{fciqmc09,i-fciqmc,sqmc}
For instance, a single-point calculation of the Chromium dimer correlating 24 electrons in 30 Hartree-Fock (HF) canonical orbitals, which would be plausibly accurate to the degree of 0.1 $mE_{\rm h}$, was reported using the initiator adaptation of FCIQMC (i-FCIQMC) with 200 million walkers spending 576 times 34 processor core hours.\cite{mp_fciqmc}
More recent advances involve the sampling of unbiased reduced density matrices,\cite{replica} nuclear gradient,\cite{grad} and complete active space self-consistent field (CASSCF) orbital optimization.\cite{casscf}
Several groups have also proposed extensions of FCIQMC for excited states.\cite{fciqmc-excit,MSQMC,Humeniuk,fciqmc-kry,fciqmc-proj}
The model space QMC (MSQMC) is one of such approaches capable of sampling formally exact quasi-degenerate and excited state wavefunctions without introducing bias, and was recently applied to extensive calculations of potential energy curves of excited states for N$_2$, O$_2$, and their ions.\cite{ohtsuka_ms}
Therefore, one of the most signifiant challenges of molecular electronic structure theory confronting us in recent years is to treat the dynamic and non-dynamic correlation effects of, {\it e.g.}, multi-nuclear transition metal complexes in a balanced manner, which requires much larger orbital space and number of interacting electrons than the limitations of currently available tools for FCI problems.
Although the full valence CAS picture has been frequently employed for systems comprising main group elements, the construction of optimum reference wavefunctions is nontrivial for transition metal complexes.

The constitution of this paper is in the following.
We first present a novel MSQMC algorithm introducing a state-selective partitioning, which requires less computational cost than the original MSQMC with the L\"owdin partitioning.
The imaginary-time evolution of the walker distribution is then expanded order-by-order in the single-reference (SR) limit for a pertrubational analysis of the stochastic wavefunction.
We further propose several size-consistency corrections to the initiator adaptation of MSQMC.
Numerical examples are presented in Sec. III followed by conclusions.

\section{THEORY}
\subsection{Multi-state MSQMC formalism}
The imaginary-time evolution (ITE) of (\ref{eq:proj}) in infinitesimal interval can be partitioned into the components, $\psi_{\rm P}(\tau)=\hat P \psi(\tau)$ and $\psi_{\rm Q}(\tau)=\hat Q \psi(\tau)$, as
\be
\frac{d}{d \tau}
\begin{pmatrix}
\psi_{\rm P}(\tau)\\
\psi_{\rm Q}(\tau)\\
\end{pmatrix}
=
-
\begin{pmatrix}
\hat P\hat H - E & \hat P \hat H \\
\hat Q \hat H & \hat Q \hat H - E
\end{pmatrix}
\begin{pmatrix}
\psi_{\rm P}\\
\psi_{\rm Q}
\end{pmatrix},\label{eq:itmc}
\ee
where $\hat P$ and $\hat Q$ are the projection operators onto the the model space (P-space) and its orthogonal component (Q-space), respectively.
MSQMC\cite{MSQMC} treats the model space amplitude $\psi_{\rm P}$ deterministically by the diagonalization of the effective Hamiltonian, and the imaginary time evolution of $\psi_{\rm Q}(\tau)$ stochastically for the fixed $\psi_{\rm P}$ as
\be
\frac { d\psi_{\rm Q}(\tau)}{d\tau} =- \hat Q (\hat H - E)\psi_{\rm Q}(\tau)-\hat Q \hat H \psi_{\rm P}.\label{eq:itetc}
\ee
Note this equation is valid not only for the ground state but also for excited states as indicated by the stationary condition $\frac { d\psi_{\rm Q}(\tau)}{d\tau}=0$.
The original MSQMC algorithm employed the energy-dependent effective Hamiltonian formalism based on the L\"owdin partitioning\cite{Loewdin1,Loewdin2} as a sufficiency condition for (\ref{eq:itetc}).
In what follows, we present more efficient approach which is closely related to the dual partitioning (DP) we discussed recently.\cite{DP}

Let us consider the matrix form of the Schr\"odinger equation,
\be
\begin{pmatrix}
{\bf H}_{\rm PP} & {\bf H}_{\rm PQ}\\
{\bf H}_{\rm QP} & {\bf H}_{\rm QQ}
\end{pmatrix}
\begin{pmatrix}
{\bf C}_{\rm PM} \\
{\bf C}_{\rm QM}
\end{pmatrix}
 = 
\begin{pmatrix}
{\bf C}_{\rm PM} \\
{\bf C}_{\rm QM}
\end{pmatrix}
{\bf \Lambda}_{\rm MM}\label{eq:schb},
\ee
where ${\bf \Lambda}_{\rm MM}$ is a diagonal matrix containing the state energies,
\be
{\bf \Lambda}_{\rm MM}=
\begin{pmatrix}
E_{1} & & 0 \\
 & \ddots & \\
0 & & E_{M}
\end{pmatrix}.
\ee
The number of solutions $M$ of our interest does not exceed the dimension of the P-space, which is usually much smaller than the size of the secondary space, $M \le N_{\rm P} \ll N_{\rm Q}$.
The dimension of the model space $N_{\rm P}$ is supposed to range from 1 to several thousands according to the desired number of solutions and the degree of quasi-degeneracy.
The P-space CI coefficients are determined from the effective secular equation in the model space,
\be
{\bf H}_{\rm PP}^{\rm eff} {\bf C}_{\rm PM} = {\bf C}_{\rm PM} {\bf \Lambda}_{\rm MM}\label{eq:schb_eff},
\ee
and the effect of the secondary space is calculated stochastically according to (\ref{eq:itetc}) or
\be
\frac{d{\bf C}_{\rm QM}(\tau)}{d\tau}=-{\bf H}_{\rm QQ}{\bf C}_{\rm QM} +{\bf C}_{\rm QM} {\bf S}_{\rm MM} -{\bf H}_{\rm QP}{\bf C}_{\rm PM},\label{eq:itm_cqm}
\ee
with a diagonal energy shift matrix ${\bf S}_{\rm MM}$.
Although the formal solution of ${\bf S}_{\rm MM}$ is ${\bf \Lambda}_{\rm MM}$, the use of the instantaneous contribution to the energies,
\be
{\bf S}_{\rm MM}(\tau)= {\rm diag} ({\bf C}_{\rm MP}^{\rm L} {\bf H}_{\rm PP}^{\rm eff}(\tau) {\bf C}_{\rm PM}),\label{eq:iinst_energy}
\ee
generally improves the convergence with the growth of walkers especially in the initial stage of ITE (\ref{eq:itm_cqm}),
where we have used the notation diag(${\bf A}$) for the diagonal matrix of ${\bf A}$, and ${\bf C}_{\rm MP}^{\rm L}$ is the matrix of left-eigen vectors with the normalization, ${\bf C}_{\rm MP}^{\rm L}{\bf C}_{\rm PM}={\bf I}_{\rm MM}$.
The effective Hamiltonian is expressed by
\be
{\bf H}_{\rm PP}^{\rm eff}={\bf H}_{\rm PP}+{\bf H}_{\rm PQ}{\bf T}_{\rm QP},\label{eq:heff}
\ee
where ${\bf T}_{\rm QP}$ is the transfer matrix to relate the CI coefficients in the P- and Q-spaces,
\be
{\bf C}_{\rm QM}(\tau)={\bf T}_{\rm QP}(\tau){\bf C}_{\rm PM}.\label{eq:tm}
\ee
${\bf H}_{\rm PP}^{\rm eff}$ of the form (\ref{eq:heff}) is a non-hermitian when several states with different energies are treated simultaneously, unlike the ${\bf H}_{\rm PP}^{\rm eff}$ in the L\"owdin partitioning.\cite{Loewdin1,Loewdin2}
Since (\ref{eq:tm}) is an underdetermined system, {\it i.e.} the unknowns of ${\bf T}_{\rm QP}$ are more than the number of conditions for ${\bf C}_{\rm QM}$, the solution is not unique unless $M = N_{\rm P}$.
It is assumed that all $M$ solutions are well-separated from Q-space.
The case $M = N_{\rm P}$ for the full determination of the model space solution corresponds to the Coope partitioning,\cite{EIP1,EIP2} that often introduces intruder state problems due to the inability of isolating all $N_{\rm P}$ states in general from the Q-space.

One way to formulate ${\bf T}_{\rm QP}$ for the reduced $M$ solutions is DP,\cite{DP} which further divides the P-space into the block A with the same dimension as $M$, and its complement of the buffer space B to avoid intruders.
Then, the transfer matrix, ${\bf T}_{\rm QP}=\begin{pmatrix} {\bf T}_{\rm QA} & {\bf T}_{\rm QB} \end{pmatrix}$ with ${\bf T}_{\rm QA}={\bf C}_{\rm QM}({\bf C}_{\rm AM})^{-1}$ and ${\bf T}_{\rm QB}=0$, suffices (\ref{eq:tm}) as far as the square matrix ${\bf C}_{\rm AM}$ is invertible.
The major drawback of DP is however the arbitrariness in the partitioning of A and B.
Alternatively, we introduce a state-selective partitioning (SSP) in this work in terms of the left eigenvector
\be
{\bf T}_{\rm QP}(\tau)={\bf C}_{\rm QM}(\tau){\bf C}_{\rm MP}^{\rm L},\label{eq:tm_mp}
\ee
as one of the solutions fulfilling (\ref{eq:tm}).
This expression is now recommended to use since ${\bf T}_{\rm QP}$ is uniquely determined without introducing a partitioning in the model space.

Using a positive integer of booster weight $n_{\rm boost}$ for the magnitude of each P-space component of the wavefunction $\psi_{\rm P}$, (\ref{eq:itm_cqm}) is operated by the death/cloning and spawning algorithms for the diagonal and off-diagonal contributions to sample walker distributions of ${\bf C}_{\rm QM}(\tau)$ for the $M$ solutions simultaneously.
The model space CI coefficients ${\bf C}_{\rm PM}$ are updated every $N_{\rm micro}$ steps.
The average of the signed occupation number on a Slater determinant over the booster weight converges to the exact FCI coefficient,
\be
\frac{\left\langle {{\bf N}_{\rm QM}(\tau)} \right\rangle _\tau}{n_{\rm boost}} ={\bf C}_{\rm QM}
\ee
under the intermediate normalization condition with respect to $\psi_{\rm P}$, where $\left\langle {A(\tau)} \right\rangle_{\tau}$ stands for the average of $A(\tau)$ over imaginary-time.

The above scheme is fast and stable if there is no intruder.
The original MSQMC\cite{MSQMC} utilizing the energy-dependent partitioning (EDP) of L\"owdin samples the transfer matrix directly for solutions in the vicinity of the target energy $E$,
\be
\frac{d{\bf T}_{\rm QP}(\tau)}{d\tau}=-({\bf H}_{\rm QQ}-E){\bf T}_{\rm QP} -{\bf H}_{\rm QP}.\label{eq:itm_eip}
\ee
EDP is more expensive requiring $N_{\rm P}$ sets of walkers for each state, in comparison with (\ref{eq:itm_cqm}) which handles the same number of walker sets as $M$.
Nevertheless, (\ref{eq:itm_eip}) is still a useful option when the target solutions contain significant degeneracies.
In the SR limit ($N_{\rm P}=1$), ${\bf T}_{\rm QP}$ reduces to ${\bf C}_{\rm QM}$, and the EDP approach coincides to MSQMC with SSP.
The main difference in the practical implementations of (\ref{eq:itm_cqm}) and (\ref{eq:itm_eip}) is the treatment of the coupling between the P- and Q-spaces.
An efficient treatment of the P-space spawning step for the coupling in SSP will be described in a separate paper.

\begin{center}
\begin{figure}[t]
\includegraphics[width=250pt]{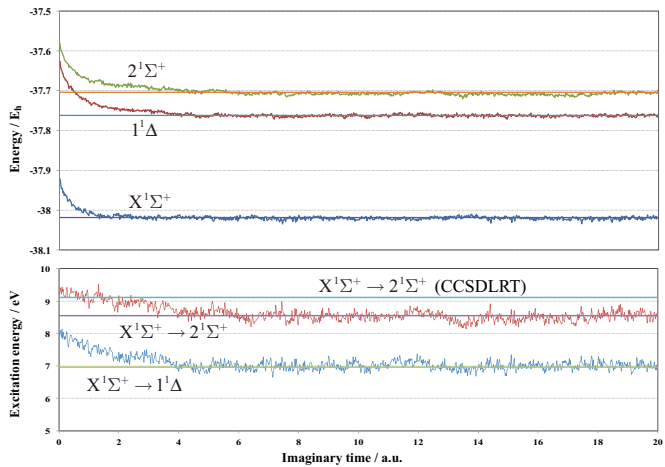}
\caption{Instantanous energies of the 3 low-lying $^1$A$_1$ electronic states in the $C_{2v}$ point group symmetry for the CH$^+$ molecule. The FCI energies\cite{Olsen_CH+} are shown in horizontal lines. The corresponding excitation energies are in the lower panel. The CCSD linear response theory (CCSDLRT) largely under-correlates the the $2^1\Sigma^+$ state, the main component of which is two-electron excitation with respect to RHF.\cite{Koch_CH+}}
\label{fig:ch+}
\end{figure}
\end{center}

Fig. \ref{fig:ch+} shows the convergence of MSQMC energies with SSP for the X$^{1}\Sigma^{+}$, 1$^{1}\Delta$, and 2$^{1}\Sigma^{+}$ states of CH$^+$ computed as $A_1$ states in the $C_{2v}$ subgroup symmetry.
The model space consists of 6 Slater-determinants out of 28 in the (4e,5o) valence complete active space using the RHF orbitals.
The MSQMC in SSP converges the 3 states to the corresponding FCI limits simultaneously unlike EDP that requires different simulations for different states.
The instantaneous energies are almost equilibrated in the imaginary-time of 5 a.u.
The P-space CI coefficients are updated every 10 a.u. imaginary-time, albeit this effect is secondary for this particular system.
EDP requires 18 computational walker sets for the 3 electronic states (6 sets for the P-space determinants times 3 states), while only 3 walker sets are used for SSP.
This computational advantage of SSP becomes particularly important when a large dimension of the model space needs to be handled for quasi-degenerated systems.

\subsection{Perturbative analysis}
The convergence of the M{\o}ller-Plesset (MP) perturbation series is investigated routinely using FCI programs.\cite{mpn_fci}
Similarly, an analysis of stochastic components can be performed perturbationally based on the MSQMC code for FCI.
Recently, Jeanmairet {\it et al.} presented an expression of ITE for a perturbative wavefunction in an arbitrary order with a multi-reference (MR) wavefunction.\cite{Jeanmairet}
The same expression can be derived readily within the framework of MSQMC, as the L\"owdin partitioning has been frequently employed to formulate perturbation theory by approximating the resolvent.\cite{Loewdin2,Freed1990,Davidson1998}

Choosing the model space wavefunction to be HF, $\psi_{\rm P}=\psi_{\rm HF}=\psi^{(0)}$, and using the standard perturbative expansion, $\hat H=\hat H_0 + \lambda \hat V$, $\psi_{\rm Q}=\sum_{i=1}^{\infty}{\lambda^i \psi^{(i)}}$, and $E=\sum_{i=0}^{\infty}{\lambda^i E^{(i)}}$, ITE (\ref{eq:itetc}) is expanded order-by-order as
\bea
\frac {d\psi^{(1)}}{d\tau} = - (\hat H_{0}&-&E^{(0)})\psi^{(1)} - \hat Q \hat V \psi^{(0)}, \\
\frac {d\psi^{(2)}}{d\tau} = - (\hat H_{0}&-&E^{(0)})\psi^{(2)} - (\hat Q \hat V - E^{(1)})\psi^{(1)}, \\
\frac {d\psi^{(n)}}{d\tau} = - (\hat H_{0}&-&E^{(0)})\psi^{(n)} - (\hat Q \hat V - E^{(1)})\psi^{(n-1)} \nonumber \\
&+& \sum_{i=2}^{n-1} E^{(i)}\psi^{(n-i)},\label{eq:it_pt}
\eea
with $E^{(0)}=\left\langle {0} \right|\hat H_0\left| {0} \right\rangle$ and $E^{(n+1)}=\left\langle {0} \right|\hat V\left| {n} \right\rangle$.
We employ the MP partitioning, $\hat H_{0} = \hat F$, along with the HF canonical orbitals.

A practical stochastic algorithm is as follows.
Representing each of the perturbed wavefunctions by a signed walker distribution, the variation of $\psi^{(n)}$ in a time step $\delta \tau$ is expressed as
\be
\delta \psi^{(n)} = \delta \psi^{(n)}_{\rm death} + \delta \psi^{(n)}_{\rm spawn} + \delta \psi^{(n)}_{\rm trans},
\ee
with
\bea
\delta \psi^{(n)}_{\rm death} &=& - \delta \tau (\hat H_{0}-E^{(0)}) \psi^{(n)}, \\
\delta \psi^{(n)}_{\rm spawn} &=& - \delta \tau \hat Q \hat H_{\rm x} \psi^{(n-1)}, \\
\delta \psi^{(n)}_{\rm trans} &=& - \delta \tau [(\hat H_{\rm d} - \hat H_{0} -E^{(1)}) \psi^{(n-1)} \nonumber \\
&+& \sum_{i=2}^{n-1} E^{(i)}\psi^{(n-i)}],
\eea
where $\hat V=\hat H_{\rm x}+\hat H_{\rm d}-\hat H_{0}$ with $\hat H_{\rm d}$ and $\hat H_{\rm x}$ standing for the Hamiltonian operators for the diagonal and off-diagonal elements in the determinantal basis, respectively.
$\delta \psi^{(n)}_{\rm death}$ is the contribution from the zeroth order Hamiltonian and energy, and $\delta \psi^{(n)}_{\rm spawn}$ is the one from the off-diagonal interaction $\hat H_{\rm x}$ with $\psi^{(n-1)}$.
$\delta \psi^{(n)}_{\rm death}$ and $\delta \psi^{(n)}_{\rm spawn}$ are calculated in a similar way of the usual death/cloning and spawning steps in the FCIQMC algorithm.\cite{fciqmc09}
$\delta \psi^{(n)}_{\rm trans}$ is a novel type from the perturbation of diagonal-type $\hat V_{\rm d} = \hat H_{\rm d}-\hat H_{\rm 0}$ along with unlinked contributions from the lower wavefunctions and energies.
We introduce the so-called {\it transcription step} to manipulate the contribution from a walker of lower wavefunctions to the same type of determinant in $\psi^{(n)}_{\rm trans}$ using the absolute value of $\delta \tau (\hat H_{\rm d} - \hat H_{0} -E^{(1)})$ or $\delta \tau E^{(i)}(\tau)$ as the probability for a transcription.
All simulations for the perturbative wavefunctions $\psi^{(n)}$ are performed in parallel using the same time step $\delta \tau$.
The population control of walkers in each order $n$ is performed only in terms of the booster weight $n_{\rm boost}$ for the HF amplitude, {\it i.e.} the number of walkers on a determinant divided by $n_{\rm boost}$ corresponds to the perturbative amplitude under the intermediate normalization with respect to HF.
Note, Thom and Alavi previously proposed a graph-based stochastic perturbation algorithm with numerical implementations through MP3\cite{qmc_mp3} and MP4.\cite{qmc_mp4}

\begin{center}
\begin{figure}[]
\includegraphics[width=250pt]{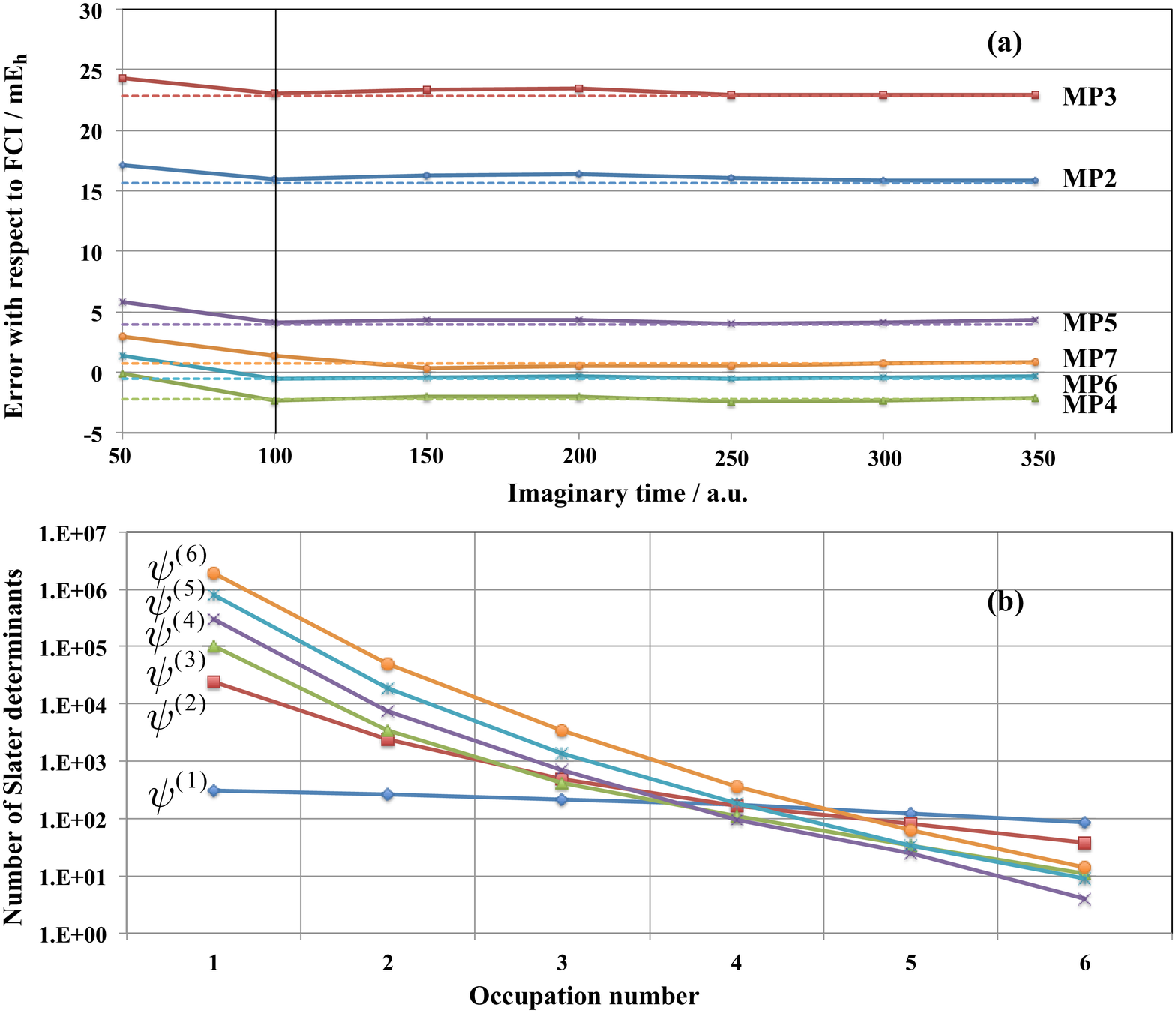}
\caption{(a) Errors of stochastic MP$n$ energies with respect to the valence electron correlated FCI result for the $X^1\Sigma_g^+$ states of N$_2$ in cc-pVDZ basis set. The FCI dimension is $5.4 \times 10^8$. The time step and booster weight for the HF determinant are $\delta \tau = 10^{-3}$ a.u. and $n_{\rm boost}=1,000$ respectively, and the integrations of perturbative energies over macro iterations started at $\tau=100$ a.u. Instantaneous energies are used for the transcriptions. The FCI energy (-109.278 340 $E_{\rm h}$) and deterministic errors in the dashed lines are taken from Ref. \onlinecite{MPn}. (b) The distribution of occupation numbers of Slater determinants in the range [1,6] for each perturbative wave function at $\tau = 350$ a.u.}
\label{fig:MPn}
\end{figure}
\end{center}

As an example, we show the errors of stochastic MP$n$ energies along with the occupation number distributions of the perturbative wave functions for N$_2$ in Fig. \ref{fig:MPn}.
All stochastic energies through MP7 converge to their deterministic limits within sub $mE_h$ in $\tau=100$ a.u.
The total number of walkers for each $\psi^{(n)}$ is dominated by singly-occupied Slater determinants except for $\psi^{(1)}$, and rapidly increases with the order of perturbation according to the excitation level $2n$ with respect to $\psi^{(0)}$.
The required number of walkers exceeds $2\times 10^6$ at $\psi^{(6)}$, and a stochastic CI or higher-order perturbation for lager systems necessitates to truncate the configuration space with small occupation numbers.
Nevertheless, any truncation in a CI expansion violates the linked diagram or Brueckner-Goldstone theorem\cite{Lindgren-Morrison} in each order,
\be
\frac { d\psi^{(n)}}{d\tau} = - (\hat H_{0}-E^{(0)})\psi^{(n)} - \hat Q (\hat V \psi^{(n-1)})_{\rm linked},
\ee
that is necessary to ensure the correct scaling of the expansion with the system size.
The linked diagram theorem was extended to MR case,\cite{Brandow} and it is likely that a general expansion in a state-specific case is also size-extensive.
Unlike the CI expansion (\ref{eq:itm_cqm}), the connectivity from the SR CC ansatz assures the linked diagram expansion,
\bea
\frac {d \hat T(\tau)}{d\tau} \psi_{\rm HF} &=&- \hat Q (\hat H \psi(\tau))_{\rm connected},\label{eq:cc_qmc1} \\
\psi(\tau) &=& e^{\hat T(\tau)}\psi_{\rm HF},\label{eq:cc_qmc2}
\eea
with the instantaneous energy $E(\tau) = \left\langle {\psi_{\rm HF}} \right| \hat H \left| {\psi (\tau)} \right\rangle$.
Especially, the stochastic simulation for its linearization,
\be
\frac {d \hat T(\tau)}{d\tau} \psi_{\rm HF} = - \hat Q (\hat H + [\hat H, \hat T(\tau)]) \psi_{\rm HF},
\ee
can be performed precisely in an arbitrary configuration space by the explicit comparison of the particle-hole indices with respect to the SR vacuum for the connected Hamiltonian matrix elements
\be
\left\langle {\mu} \right|\hat H \left| {\nu} \right\rangle_{\rm connected} = \left\langle { \hat \mu^{\dagger} 0} \right| [\hat H,\hat \nu] \left| {0} \right\rangle,
\ee
where $\mu^{\dagger}$ and $\nu^{\dagger}$ denote excitation operators, and $\left| \mu \right\rangle = \hat \mu^{\dagger} \left| 0 \right\rangle$ and $\left| \nu \right\rangle = \hat \nu^{\dagger} \left| 0 \right\rangle$.
This feature was also recently discussed by Thom and coworkers in their linked version\cite{lccmc} of the stochastic CC approach.\cite{ccmc}
(\ref{eq:cc_qmc1}) and (\ref{eq:cc_qmc2}) become the formally exact CC only in the thermodynamic limit, {\it i.e.} $n_{\rm boost} \to \infty$ unlike the linearized case as one can easily gather from the contrary extreme of $\hat T(\tau)$ represented by very small number of walkers.
In the following, we present more pragmatic approaches to remedy the size-inconsistency error of stochastic CI with more general reference spaces.

\begin{center}
\begin{figure}[t]
\includegraphics[width=200pt]{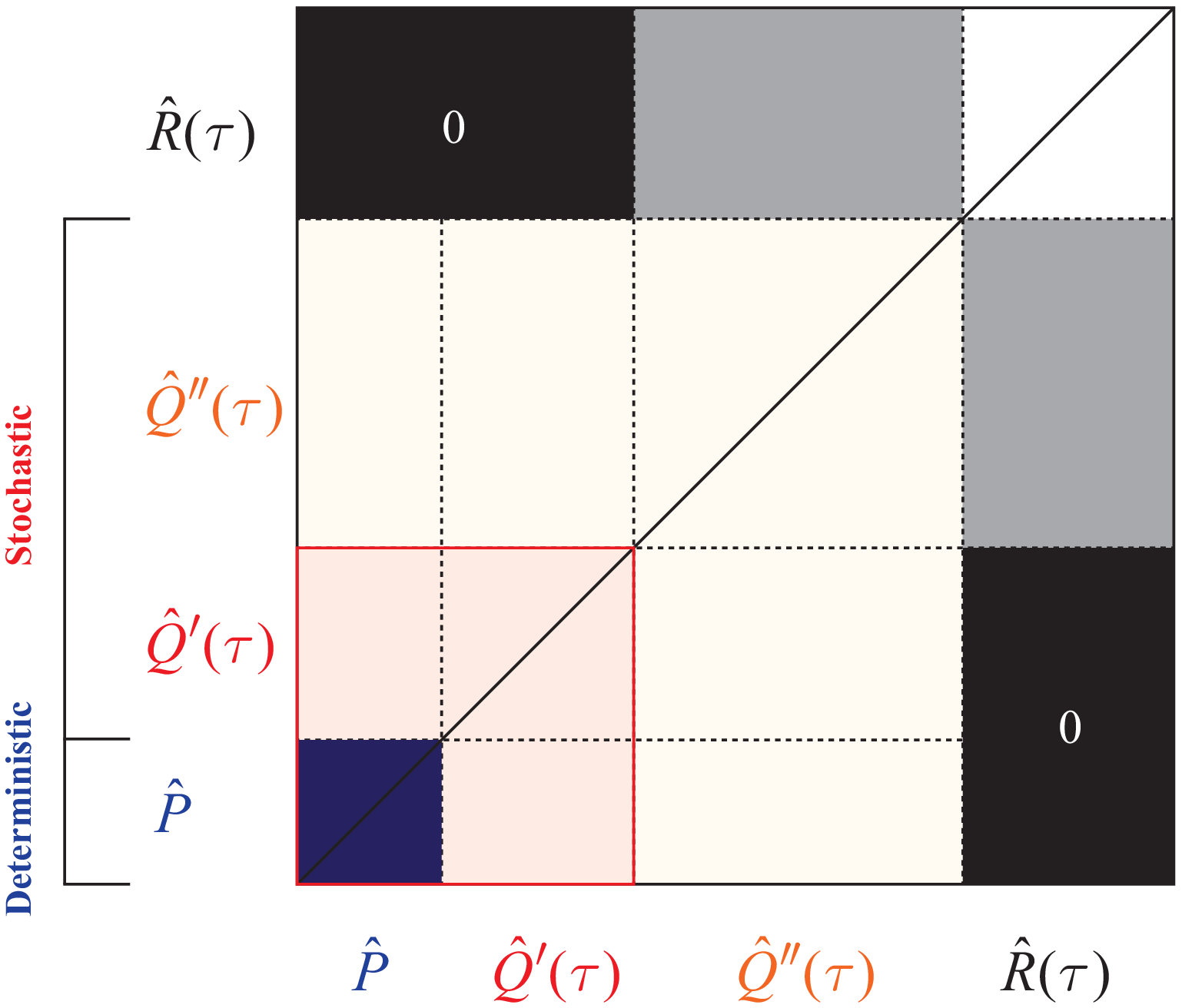}
\caption{Definitions of the configuration spaces of MSQMC in a truncated configuration space. The reference space consists of $\hat P$ and $\hat Q'$ (in the red square), and the first order interacting space $\hat Q''$ defines the stochastic boundary.}
\label{fig:config}
\end{figure}
\end{center}

\subsection{Size-consistency corrections}
Let us consider the interacting spaces as depicted in Fig. \ref{fig:config}.
The stochastic space which can be dependent on $\tau$ is divided into the primary and secondary parts,
\be
\hat Q(\tau)=\hat Q'(\tau)+\hat Q''(\tau).
\ee
The secondary part of the projector $\hat Q''(\tau)$ is the first order interacting space of the union of the fixed deterministic and primary stochastic ones, $\hat P + \hat Q'(\tau)$, defined as the reference space.
The use of all higher-excitations in $\hat R(\tau)$ becomes impractical for large molecules due to the growth of the number of walkers especially on singly occupied Slater-determinants.
In other words, we attempt to build a size-consistent model based on a stochastic MRCI with the flexible reference space, $\hat P + \hat Q'(\tau)$, approximately taking account of the contribution from the outer space $\hat R(\tau)$ containing the effects of triples and quadruples with respect to the reference space.
Variations in the interaction and imaginary-time regimes lead to several options to construct the configuration spaces.
The initiator approach\cite{i-fciqmc} restricts the configuration space by the interaction with initiator determinants possessing population exceeding $T_{\rm I}$, and can be deemed as such a dynamical MRCI variant for the projectors $\hat Q' (\tau)$ and $\hat Q''(\tau)$ spanned by the initiator and non-initiator determinants, respectively.
We have implemented the initiator adaptation of MSQMC (i-MSQMC) in two ways.
One is based on {\it the survival criterion} of Cleland {\it et al.},\cite{i-fciqmc} which choose a progeny from all determinants interacting with a parental walker.
When merging the lists of occupied determinants and newly spawned walkers, a progeny deriving from noninitiator is killed once it turns out to be on an unoccupied determinant.
The survival criterion necessities the manipulation of the walkers in $\hat R$ (triples and quadruples with respect to initiator), most of which do not live on finally.
The other implementation, which gives an equivalent result as the survival criterion, employs {\it a birth selection} for the spawning of the noninitiator determinants, {\it i.e.}, candidates attempted to be spawned from noninitiator are selected from already occupied determinants regardless of the connectivity to the parent with a uniform probability $1/N_{d}$, $N_{d}$ being the number of Slater determinants in the list of occupied walkers.
If a selected progeny coincides to the parental determinant, the spawning event does not take place.
This simple list spawning (LS) algorithm enables the uniform sampling in the stochastic space $\hat Q(\tau)$ very quickly, albeit a small time step is sometimes needed due to the increase of noninteracting determinants in the walker list with a large number of initiators.
The initiator approach restrains the configuration space as a truncated CI, and we need to take the correct scaling of the correlation energy into account as numerically observed previously.\cite{ohtsuka_ms}

\subsubsection{Coupled electron pair approximations}
We first consider the static limit of the configuration spaces.
The ITE of the CI coefficients in the primary stochastic space ${\bf C}_{\rm Q'M}$ coincides to (\ref{eq:itm_cqm}),
\bea
\frac{d{\bf C}_{\rm Q'M}(\tau)}{d\tau}=&-&{\bf H}_{\rm Q'Q}{\bf C}_{\rm QM} +{\bf C}_{\rm Q'M} {\bf S}_{\rm MM} \nonumber \\
&-&{\bf H}_{\rm Q'P}{\bf C}_{\rm PM},\label{eq:itm_cq'm}
\eea
while there exists a coupling with $\left| R \right\rangle$ for ${\bf C}_{\rm Q''M}(\tau)$,
\bea
\frac{d{\bf C}_{\rm Q''M}(\tau)}{d\tau}=-{\bf H}_{\rm Q''Q}{\bf C}_{\rm QM} +{\bf C}_{\rm Q''M} {\bf S}_{\rm MM} \nonumber \\
-{\bf H}_{\rm Q''P}{\bf C}_{\rm PM}-{\bf H}_{\rm Q''R}{\bf C}_{\rm RM} .\label{eq:itm_cq''m}
\eea
The L\"owdin partitioning relates ${\bf C}_{\rm RM}$ and ${\bf C}_{\rm Q''M}$ explicitly as
\be
{\bf H}_{\rm RQ''}{\bf C}_{\rm Q''M}+{\bf H}_{\rm RR}{\bf C}_{\rm RM}-{\bf C}_{\rm RM}{\bf \Lambda}_{\rm MM}=0,
\ee
and yet the treatment of the enormous objects with R is impractical.
Instead of explicitly using ${\bf C}_{\rm RM}$, the coupling is represented in a way analogous to the usual MR coupled electron-pair approximations (CEPA) as
\be
-{\bf H}_{\rm Q''R}{\bf C}_{\rm RM} \approx - a {\bf C}_{\rm Q''M} {\bf \Lambda}''_{\rm MM},
\ee
where ${\bf \Lambda}''_{\rm MM}$ is the correlation energy contribution from the secondary space,
\bea
{\bf \Lambda}''_{\rm MM}&=& {\rm diag} ({\bf C}_{\rm MP}^{\rm L} {\bf V}_{\rm PP}^{\rm eff} {\bf C}_{\rm PM}), \\
{\bf V}_{\rm PP}^{\rm eff}&=&{\bf H}_{\rm PQ''}{\bf T}_{\rm Q''P}.
\eea
The parameter $a$ takes $a_{0}=1$, $a_{\rm P}=1-\frac{2}{N_{\rm e}}$, and $a_{\rm M}=\frac{(N_{\rm e}-2)(N_{\rm e}-3)}{N_{\rm e}(N_{\rm e}-1)}$ for linearized coupled-cluster (LCC) or CEPA0 class of methods,\cite{lcc1,lcc2,Cave,Tanaka} averaged coupled pair functional (ACPF),\cite{acpf} averaged quadratic coupled-cluster (AQCC),\cite{aqcc} respectively.
For a survey of approximately size-consistent modifications of MRCI, readers can refer to Ref. \onlinecite{cpall}.
Consequently, the ITE (\ref{eq:itm_cq''m}) becomes
\bea
\frac{d{\bf C}_{\rm Q''M}(\tau)}{d\tau}=&-&{\bf H}_{\rm Q''Q}{\bf C}_{\rm QM} +{\bf C}_{\rm Q''M} {\bf S}_{\rm MM}^{(0)} \nonumber \\
&-&{\bf H}_{\rm Q''P}{\bf C}_{\rm PM},\label{eq:itm_cq''m_cepa0}
\eea
with the new shift of the instantaneous energy excluding the $Q''$-space contribution,
\be
{\bf S}_{\rm MM}^{(0)}(\tau)= {\rm diag} [{\bf C}_{\rm MP}^{\rm L} ({\bf H}_{\rm PP}^{\rm eff}(\tau)-a{\bf V}_{\rm PP}^{\rm eff}(\tau)) {\bf C}_{\rm PM}],\label{eq:shift_cepa0}
\ee
and the different shifts, ${\bf S}_{\rm MM}(\tau)$ and ${\bf S}_{\rm MM}^{(0)}(\tau)$ are used for the initiator and noninitiator determinants, respectively, for the CEPA corrections.
In that sense, the entire framework of the present approach (besides the choice of $a$) bears a closer resemblance to quasi-degenerate variational perturbation theory (QDVPT)\cite{Cave} using effective Hamiltonian to account for the relaxation of the reference wavefunction, rather than the LCC methods.\cite{lcc1,lcc2}
We shall dub, {\it e.g.}, the linearized coupled-cluster singles-and-doubles (CCSD) variant of i-MSQMC as i-CEPA0-MSQMC, which reduces to the stochastic CEPA0 in the SR limit, {\it i.e.}, $\hat Q'=0$ with $\psi_{\rm P}=\psi_{\rm HF}$, owing to the Brillouin condition if the stochastic space is sampled uniformly.
The non-initiator contribution to the instantaneous correlation energy ${\bf V}_{\rm PP}^{\rm eff}(\tau)$ decreases with improving the primary space energy, and the the CEPA and a posteriori corrections ({\it vide infra}) vanish as the number of walkers goes infinity with a fixed $T_{\rm I}$.
Jeanmairet {\it et al.}\cite{Jeanmairet} recently developed a stochastic LCC method in a different framework with an active space using the partitioning of Fink\cite{Fink1,Fink2} along with multi-replica samplings for the CAS reference and correlated wavefunctions.
Indeed MR-CEPA0-type approximations were derived perturbationally by Cave and Davidson,\cite{Cave} and Tanaka {\it et al.}\cite{Tanaka} earlier in the late 1980s.
In the approach of Jeanmairet {\it et al.}, the initiator plays a role to control the quality of the reference wavefunction in CAS with different asymptotic limit with respect to the initiator threshold, in contrast to the FCI asymptotic limit of the present method.
Although there exits more rigorous CEPA\cite{Debashis_cepa} based on state-specific MRCC,\cite{Debashis_sscc} the stochastic adaptation of the model including the treatment of somewhat complex operator coupling along with the exclusion principle violating terms is beyond the scope of this work.

\subsubsection{A posteriori corrections}
We can alternatively introduce a posteriori quadruple corrections using  ${\bf \Lambda}''_{\rm MM}$ of genuine i-MSQMC,
\be
E_{\kappa}^{\rm (+Q)}= \frac{a\omega''_{\kappa}}{1+\omega'_{\kappa}}\Lambda''_{\kappa \kappa},\\
\ee
where $\omega'_{\kappa}=||\left\langle {{\bf C}_{{\rm Q'}\kappa}(\tau)} \right\rangle_{\tau}||$ and $\omega''_{\kappa}=||\left\langle {{\bf C}_{{\rm Q''}\kappa}(\tau)} \right\rangle_{\tau}||$ are the weights of the $Q'$- and $Q''$ contributions under the intermediate normalization with respect to $\psi_{\rm P}$, and the parameters $a_{0}$, $a_{\rm P}$, and $a_{\rm M}$ give the Bruckner (or renormalized Davidson),\cite{BC} Pople,\cite{PC} and Meissner\cite{MC} corrections, as perturbative approximations of CEPA0, ACPF, and AQCC, respectively.
The average weights should be calculated from incoherent walker distributions as those in the replica ensembles of walkers.\cite{replica}
We estimate these quantities retaining the computational cost of a single ensemble of MSQMC, using the walker distributions at different imaginary-time.
The distribution ${\bf N}_{\rm QM}(\tau)$ is recorded at $\Delta\tau$ imaginary-time intervals, as  ${\bf N}_{\rm QM}(\tau_1)$, and the previous ${\bf N}_{\rm QM}(\tau_1)$ is moved to ${\bf N}_{\rm QM}(\tau_2)$ at the same time.
The weights are averaged using the population at $\tau_2$ as $\omega'_{\kappa}=\left\langle {{\bf C}_{{\rm Q'}\kappa}^{+}(\tau) {\bf C}_{{\rm Q}\kappa}(\tau_2)} \right\rangle_{\tau}$ and $\omega''_{\kappa}=\left\langle {{\bf C}_{{\rm Q''}\kappa}^{+}(\tau) {\bf C}_{{\rm Q}\kappa}(\tau_2)} \right\rangle_{\tau}$periodically as ${\bf N}_{\rm QM}(\tau_2)$ is refreshed.

\subsubsection{Second order perturbative corrections}
A more drastic approximation is the use of second order corrections replacing the Hamiltonian matrix elements in ${\bf{H}_{\rm Q''Q''}}$ by the zeroth-order ones in perturbation theory,
\be
\left\langle {\mu''} \right|\hat H -E \left| {\nu''} \right\rangle \Rightarrow \left\langle {\mu''} \right|\hat H_0 -E^{(0)} \left| {\nu''} \right\rangle,
\ee
where $\mu''$ and $\nu''$ denote Slater determinants in $Q''$. 
Since there is no unlinked contribution in the second order energy, a suitable choice of the zeroth order Hamiltonian and interaction spaces may provide an attractive many-body alternative to the size-consistency corrections.
Despite various choices of $\hat H_{0}$ for multireference wavefunctions, we examine only the second-order Epstein-Nesbet (EN2) correction, $\left\langle {\mu''} \right|\hat H_0\left| {\nu''} \right\rangle=\delta _{\mu'' \nu''}\left\langle {\mu''} \right|\hat H\left| {\mu''} \right\rangle$, simply omitting the spawning from the $Q''$-space (noninitiator) Slater determinants and using the same shift for i-CEPA0-MSQMC.
The EN2 correction will not be used for further discussions since i-EN2-MSQMC appears to be far inferior to the other size-consistency corrections as provided in the supplementary material.
This does not contradict to the excellent results of EN2 corrections based on selected CI reported by several groups recently.\cite{Evangelista2016,Sharma2017,Caffarel2017}
(See also the recent developments for efficient elected CI algorithms.\cite{Holmes2016,Zimmerman2017})
The number of determinants employed in these works is in the degree of $10^3-10^7$, that is much greater than the number of the initiator determinants employed in this work typically of $10^0-10^3$.
The perturbative corrections only approximates the amplitudes in singles and doubles with respect to the primary space, and no additional contributions from higher excitations is generally expected unlike the CEPA and a posteriori corrections.
It is known that the EN partitioning does not perform excellently for size-consistency,\cite{MP-EN} and other choices of $\hat H_{0}$ may be left for investigations in future.

\section{Numerical examples}
\subsection{Size-consistency for the non-interacting Ne$_2$}
\begin{center}
\begin{figure}[t]
\includegraphics[width=250pt]{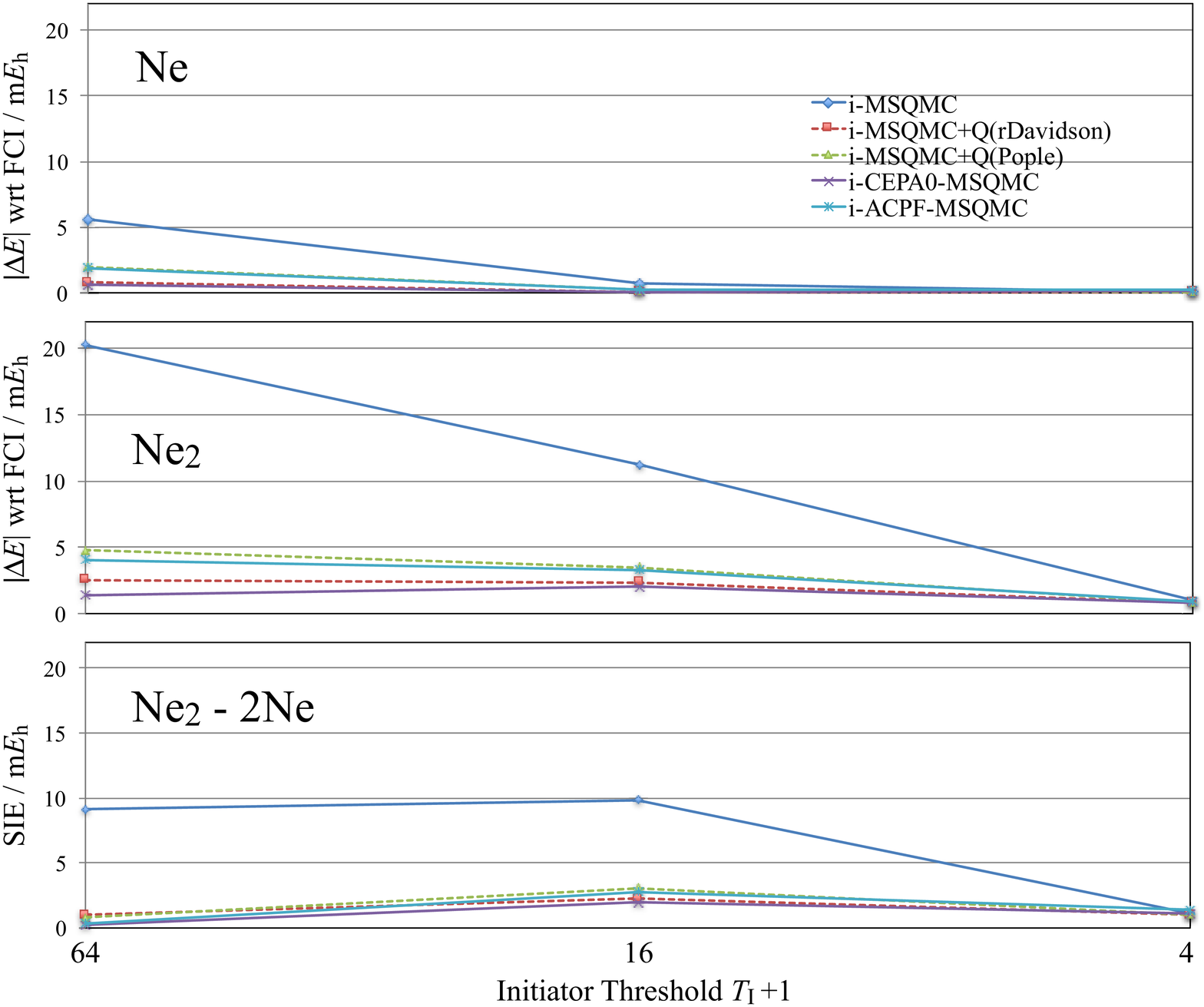}
\caption{Errors of total energies for Ne and noninteracting Ne$_2$, and size-inconsistency error with respect to FCI using different $T_{\rm I}$.}
\label{fig:sc}
\end{figure}
\end{center}

We examine the ability of the corrections of i-MSQMC in reducing the size-inconsistency error (SIE) using Ne.
Fig. \ref{fig:sc} shows the error of i-MSQMC energies for Ne and Ne$_2$ along with the size-inconsistency error with respect to FCI in the cc-pVDZ basis set by changing the initiator threshold with $n_{\rm boost}=1,000$.
The original state energies are provided in the supplementary material.
At the largest initiator threshold, $T_{\rm I}+1=64$, the RHF determinant constituting the model space is the only initiator to define the SD interacting space.
Accordingly, i-MSQMC with this $T_{\rm I}$ corresponds to the stochastic counterpart of CI singles-and-doubles (CISD), the energies of which, -128.6734(2) and -257.3378(1) $E_{\rm h}$, for Ne and Ne$_2$ are in agreement with CISD, -128.673617 and -257.338282 $E_{\rm h}$, respectively, notwithstanding that the stochastic methods do not sample the full SD space uniformly but biased by the instantaneously space occupied by walkers.
In this case, SIE of i-MSQMC amounts to 9 $mE_{\rm h}$.
Similarly, i-CEPA0-MSQMC with $T_{\rm I}+1=64$ is the the stochastic counterpart of size-extensive CEPA0.
The i-CEPA0-MSQMC energies are -128.6784(2) and -257.3567(2)$E_{\rm h}$ for Ne and Ne$_2$, which compare well to the deterministic CEPA0 energies, -128.678603 and -257.357206 $E_{\rm h}$, exhibiting almost perfect size-consistency.
SIE is also negligibly small for i-ACPF-MSQMC, and is slightly increased for i-MSQMC with each of the a posteriori corrections.

We turn to the convergence with respect to the initiator threshold.
For the monomer, all i-MSQMC energies converge to the FCI limit, -128.679025 $E_{\rm h}$, very rapidly with tightening the initiator threshold, and are in agreement with the limit within the stochastic errors at $T_{\rm I}+1=4$.
Contrarily, the convergence for Ne$_2$ is much more deteriorated.
The error of i-MSQMC at $T_{\rm I}+1=16$ is more than 10 $mE_{\rm h}$, the amount of which is almost equivalent to the size-inconsistency error due to the unbalanced descriptions between 2Ne and Ne$_2$.
SIE is only slightly increased with this $T_{\rm I}$ compared to $T_{\rm I}+1=4$ in all cases.
It is considered that this large error is originating from the structure of the initiator space for the non-intercting Ne dimer, which does not span the direct product of those for monomers.
The errors of the size-consistency corrections are insensitive to $T_{\rm I}$ for the dimer, and almost coincide to the i-MSQMC one at $T_{\rm I}+1=4$ with residual errors ca. 1$mE_{\rm h}$.
In the above range of $T_{\rm I}$, the number of walkers increases from ca. $7\times 10^3$ to $3\times 10^4$.
Nevertheless, the convergence beyond this regime, in which the size-consistency corrections become negligibly small, is painfully slow to require 1-2 order of magnitude large number of walkers by increasing $n_{\rm boost}$ to reduce the residual error to 0.1 $mE_{\rm h}$ even for this small system.

\subsection{Excited states of C$_2$}
\begin{center}
\begin{figure}[t]
\includegraphics[width=250pt]{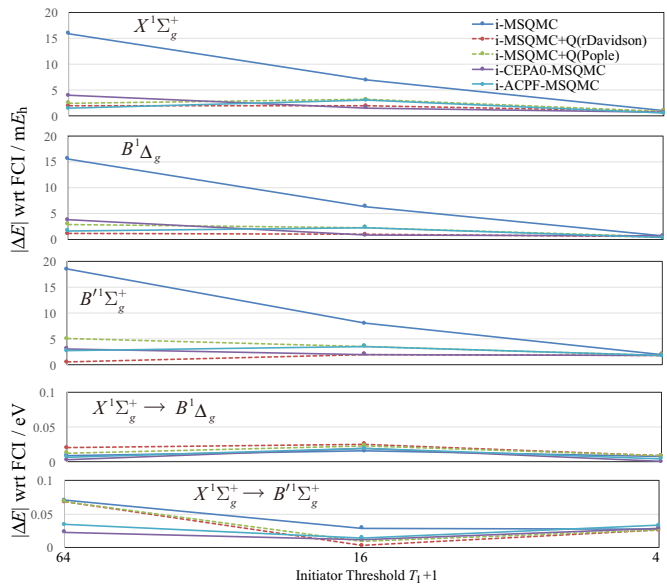}
\caption{Absolute errors of the state (upper 3 panels) and excitation (lower 2 panels) energies for $X^{1}\Sigma_{g}^{+}$, $B^{1}\Delta_{g}$, and $B'^{1}\Sigma_{g}^{+}$ of C$_{2}$.}
\label{fig:c2}
\end{figure}
\end{center}

We show the effectiveness of the size-consistency corrections holds for excited states.
We calculate the $X^{1}\Sigma_{g}^{+}$, $B^{1}\Delta_{g}$, and $B'^{1}\Sigma_{g}^{+}$ states of C$_{2}$, all of which belong to $^1A_g$ in the $D_{2h}$ computing subgroup symmetry.
The FCI result\cite{c2_fci} in the 6-31G* set is used as the reference at $R$(C-C)=1.25 {\AA}.
The model space is constructed by distributing 6 electrons in 4 RHF orbitals excluding $1s$ core and $2s$ $\sigma_g$ orbitals, and 4 determinants most dominating the 3 electronic states in the CASCI are selected.
Fig. \ref{fig:c2} shows the errors of the calculated state and excitation energies with respect to FCI.
The state energies of the simulations are also provided in the supporting material.
Similarly to the behavior in the size-consistency examination, the large error of i-MSQMC energy at $T_{\rm I}+1=64$ monotonically decreases with tightning $T_{\rm I}$.
In contrast, all errors of i-MSQMC with size-consistency corrections are quite small and uniformly distributed throughout the range of $T_{\rm I}$.
The effects of the size-consistency corrections almost disappear at $T_{\rm I}+1=4$.
Among the 3 states, the convergence for $B'^{1}\Sigma_{g}^{+}$ is somewhat slower.
A relatively large error ca. 1 $mE_{\rm h}$ appears to remain only for this state even with the increased booster weights $n_{\rm boost}=2,000$.
It is likely that this is partially due to the use of the RHF orbitals optimized solely with respect to the ground state.
For excitation energies, the majority of the i-MSQMC error tends to cancel, albeit the i-CEPA0- and i-ACPF-MSQMC clearly outperform i-MSQMC and those with a posteriori corrections for the $X^{1}\Sigma_{g}^{+} \to B'^{1}\Sigma_{g}^{+}$ at $T_{\rm I}+1=64$.
Overall, the renormalized Davidson correction (the perturbative counterpart of CEPA0) performs very similarly to i-CEPA0-MSQMC when the correction is small enough for $T_{\rm I}+1 \le 16$.

\subsection{The chromium dimer}
The chromium dimer has been used as a benchmark system to assess methods for strongly correlated systems.
Booth and coworkers\cite{mp_fciqmc} performed a state-of-the-art i-FCIQMC calculation for this system at a bond length of 1.5 {\AA} correlating 24 electrons in 30 RHF orbitals of the SV basis set.\cite{sv}
The SV basis set is too small for obtaining a quantitatively accurate dissociation energy and spectroscopic constants, yet is a suitable choice for the initial application of the present size-consistency corrections, as the essential character of the non-dynamic correlation effects of Cr$_2$ around the equilibrium distance, {\it i.e.} the competing bonding of 4$s$ and 3$d$ and repulsion of 3$p$ orbitals, is subsumed in the model.

In TABLE \ref{tab:cr2}, we show the energy and number of walkers of the i-MSQMC methods using the single determinant model space of RHF along with $n_{\rm boost}=200$ and $T_{\rm I}=9$.
The stochastic errors were estimated from the standard error of 3 sets of i-MSQMC simulations for 500 a.u. imaginary-time.
The number of walkers $N_{\rm w}$ employed in the i-MSMQC calculation under these conditions is only less than ten-thousandth of that of the most advanced i-FCIQMC calculation of Booth {\it et al.}
Accordingly, the error of the i-FCIQMC energy is ca. 64 $mE_{\rm h}$.
Actually, the convergence of the i-MSQMC energy is very slow and eventually approaches to the FCI limit with increasing the order of $N_{\rm w}$ by changing $T_{\rm I}$ or $N_{\rm boost}$ to the i-FCIQMC magnitude.
Contrarily, the size-consistent corrections are quite effective, {\it i.e.} the renormalized Davidson correction recovers more than 90\% of the error of i-MSQMC at the same computational cost, and i-CEPA0-MSQMC is accurate to sub-milli $E_{\rm h}$.
The number of walkers in i-CEPA0-MSQMC is somewhat larger than the i-MSQMC one due to the increase in the shift ${\bf S}^{(0)}_{\rm MM}(\tau)$ to prolonging the lifetime of walkers.
Note it appears that this result includes some error cancelation, and increasing $n_{\rm boost}$ does not necessarily improve the results of the size-consistency corrections systematically.
A larger booster weight, $n_{\rm boost}=400$, with the same $T_{\rm I}$ leads to the i-MSQMC energy, -2086.3790(2) $E_{\rm h}$, and the i-CEPA0-MSQMC one, 2086.407(1) $E_{\rm h}$, deteriorated by more than 10 $E_{\rm h}$ compared to the case with $n_{\rm boost}=200$.
This is because the increase of $n_{\rm boost}$ reduces proportion of the occupied determinants to those in the first order interacting space of the initiators in the survival criterion for the initiator approximation.
At any rate, the present result is encouraging strongly indicating that the size-consistent corrections combined with the initiator approach are promising means to treat strongly-correlated systems.
A further investigations with wider range of applications will be reported in a separate paper.

\begin{table}
\caption
{\label{tab:cr2}
Total energy and number of walkers of i-MSQMC simulations for the chromium dimer.}
\begin{tabular}{lccc}
\hline
& Method & Energy ($E_{\rm h}$) & $N_{\rm w}$\\
\hline
& i-MSQMC & -2086.3573(3) & $1.5 \times 10^4$ \\
& i-MSQMC+Q(rDavidson) & -2086.4154(2) & $1.5 \times 10^4$ \\
& i-CEPA0-MSQMC & -2086.4207(2) & $2.3 \times 10^4$ \\
& i-FCIQMC\footnote[1]{Near FCI result of Booth {\it et al}.\cite{mp_fciqmc}} & -2086.4212(3) & $2.0 \times 10^8$ \\
\hline
\end{tabular}
\end{table}

\section{CONCLUSIONS}
The advancements made in this paper are as follows.
(i) We introduced the state-selective partitioning (SSP) into MSQMC, that enables us to compute multi-electronic states simultaneously at low computational cost compared to the original MSQMC with the L\"owdin partitioning.
(ii) We then perturbationally analyzed the MSQMC wave function using the the MP partitioning without truncation in the configuration space.
It was shown that the number of Slater determinants with small population in $\psi^{(n)}$ increases with the order of the perturbation $n$ to discuss retaining the size-extensivity in the Hilbert space sampling.
(iii) For more general applications, we introduced CEPA and a posteriori corrections of i-MSQMC on the basis of the dynamical MRSDCI construction of the configuration spaces in the  the initiator approach.
It was numerically shown that most of the initiator error is arising from the deficiency in proper scaling with the system size, that can be greatly mitigated by the size-consistency corrections of i-MSQMC when a large initiator threshold is employed, as likely happens in most of interesting applications.
It is considered that the size-consistency corrections are effective when the amplitudes of higher-order excitations exceeding quadruple with respect to the primary space are sufficiently small.
And thus a proper choice of orbitals would be important especially for their applications to excited states and strongly correlated systems.
Finally, we suggest deterministic and semi-deterministic counterparts of the present approach using selected configuration spaces will be fast alternatives to the purely stochastic implementation.
We shall report on more comprehensive applications containing such aspects possibly combined with explicitly correlated methods\cite{F12perspective} in future.

\section{SUPPLEMENTARY MATERIAL}
See supplementary material for the detailed data used for Figs. \ref{fig:sc} and \ref{fig:c2}.

\begin{acknowledgements}
This research was partly supported by MEXT as "Priority Issue on Post-K computer" (Development of new fundamental technologies for high-efficiency energy creation, conversion/storage and use). 
This paper is dedicated to the memory of my beloved daughter, Lenka Ten-no.
\end{acknowledgements}

\end{document}